# Operando Surface Probe Microscopy Reveals Electrochemical Origins of Reliability Variability in Hafnia Ferroelectrics

**Anudeep Tullibilli*, Kartick Biswas, Shubham Kumar Parate, and Pavan Nukala***

Centre for Nano Science and Engineering, Indian Institute of Science, Bengaluru 560012, India

**ABSTRACT:**

Device-to-device and sample-to-sample variability remains a major obstacle to the reliable implementation of hafnia-based ferroelectrics, even in nominally identical structures fabricated under similar processing conditions. This highlights the need for operando diagnostic probes that provide process-relevant feedback beyond conventional structural phase analysis. Here, we demonstrate that operando nanoscale surface measurements serve as a sensitive probe of the local electrochemical processes governing reliability in hafnia ferroelectric devices. Using epitaxial rhombohedral $Y{:}HfO_2$ films grown on $La_{0.67}Sr_{0.33}MnO_3$ (LSMO)-buffered $SrTiO_3$ as a model system, we combine operando atomic force microscopy, conductive AFM, piezoresponse force microscopy, and correlative STEM-EELS to directly link surface evolution with interfacial electrochemistry. Devices exhibiting wake-up show a spatially uniform increase in surface height, arising from homogeneous oxidation of the LSMO electrode followed by oxygen migration into the ferroelectric layer during polarization switching. In contrast, leaky devices display pronounced surface roughening and blister formation, originating from spatially inhomogeneous oxidation states in the pristine LSMO electrode that promote competing oxygen evolution reactions at the LSMO/$Y{:}HfO_2$ interface. The accompanying electron generation produces leakage which is transient. Guided by these mechanistic insights, we introduce simple modifications to the processing conditions, without altering the optimized ferroelectric phase, that suppress the competing electrochemical pathways and reproducibly yield 100% ferroelectric devices. More broadly, this work establishes operando surface evolution as a powerful nanoscale diagnostic of hidden electrochemical processes and provides a practical route for improving the reliability and reproducibility of hafnia-based ferroelectric devices.

## Introduction

Hafnium oxide–based thin films have emerged as one of the most promising ferroelectric material systems for next-generation electronic devices due to their excellent compatibility with complementary metal–oxide–semiconductor (CMOS) technology and scalability to nanometer thicknesses[1,2]. In contrast to conventional perovskite ferroelectrics, whose ferroelectric properties deteriorate with decreasing thickness owing to depolarization fields and interfacial effects, hafnia-based ferroelectrics retain robust ferroelectricity even at thicknesses of only a few nanometers[2–4]. This intrinsic scalability overcomes a long-standing limitation that has hindered the integration of traditional ferroelectrics into advanced CMOS nodes[2,4,5]. Notably, doped hafnia systems with optimized compositions can exhibit remanent polarization values exceeding 20μC $cm^{-2}$ in films thinner than 10 nm, marking a significant advancement in the development of scalable ferroelectric devices[6–8].

Hafnium oxide exhibits several structural polymorphs, including cubic ($Fm\bar{3}m$), monoclinic ($P2_1/c$), tetragonal ($P4_2/nmc$), multiple orthorhombic ($Pca2_1$, $Pmn2_1$, $Pbca$) and rhombohedral phases ($R3$, $R3m$)[4,8–11]. Among these, metastable phases such as O- ($Pca2_1$), $Pmn2_1$, $R3m$ and $R3$ phases show a unique polar axis that sustain ferroelectricity[8–10,12]. These phases can be stabilized through cation doping, epitaxial strain, stress engineering, controlled nano structuring, oxygen vacancy formation, or interfacial effects [8,9,13–17]. In this context, hafnia ($HfO_2$) thin films have been extensively explored for ferroelectric applications, as thin-film processing enables access to these metastable phases[2,16,17]. Early studies primarily focused on polycrystalline films grown by atomic layer deposition (ALD) followed by rapid thermal annealing, where confinement and capping layers played a key role in stabilizing the orthorhombic ferroelectric phase[2,4,17,18]. RF magnetron sputtering and ion-beam methods have also been successfully employed to grow epitaxial Y-, Fe-, and Ce-doped $HfO_2$ films, often combined with post-deposition annealing to induce ferroelectricity[19–21]. For epitaxial films, pulsed laser deposition (PLD) has been the most widely used technique, with demonstrated control over phase stabilization and crystalline orientation[8,10,11,22–24].

Irrespective of the processing technique, it is by now established that oxygen vacancies and their dynamics play a key role in stabilizing the polar phase[11,25–28] and ferroelectric switching. However, they can also act as the dominant defects responsible for the wake-up effect and leakage in $HfO_2$-based films, both of which are undesired for ferroelectric capacitors[14,28–32]. As a result, a delicate balance exists between a vacancy concentration required to sustain ferroelectricity and defect densities and configurations that promote leakage currents[14,31,32]. Notably, different devices fabricated from nominally identical polar hafnia-based films in terms of crystal structure can exhibit either robust ferroelectric switching or dominant leakage currents. This is commonly observed yet rarely reported phenomenon.

On the other hand, wake-up effect, characterized by an enhancement of ferroelectric response upon repeated electric-field cycling, has been widely reported and is commonly associated with changes in defect configurations, internal fields, or interfacial conditions[28,30–32]. Although epitaxial hafnia films grown by PLD generally show minimal wake-up behavior, a few initial switching cycles are sometimes necessary to achieve the stabilized remanent polarization[8,10,22].

Recent operando microscopy studies have shown that ferroelectric switching in hafnia-based devices is strongly coupled with oxygen vacancy redistribution, electrochemical reactions and associated structural phase transitions under electrical bias[25,33–35]. These structural transformations between various phases are accompanied by local volume changes and lattice distortions, which can generate surface morphological modifications[25,34,36]. Conversely probing the morphological evolution using local surface probe techniques under bias, can give information about underlying electrochemistry, phase transitions, and shed light on device to device, and sample to sample behavior variations beyond atomic structural differences.

In this work we show that monitoring the evolution of surface roughness and topography under applied bias is a sensitive probe of defect dynamics that govern leakage currents and wake-up behavior. By combining complementary techniques such as AFM, conductive AFM (c-AFM), and piezoresponse force microscopy (PFM), we correlate morphological modifications with local conductivity and ferroelectric switching, enabling a comparative investigation of ferroelectric and leaky regions within the same material system. In addition, electron energy-loss spectroscopy (EELS) was employed to determine the electrochemistry and oxidation state evolution responsible for differential device behavior in different regions. Together, these complementary local and spectroscopic probes provide significant nanoscale insights into the defect-mediated mechanisms underlying device-to-device variations in hafnia-based ferroelectric films. As they directly probe the local defect and electronic landscape, they provide more actionable feedback for optimizing device performance and reliability.

## Results

### Structural, Morphological and Electrical Characterization

Epitaxial YHO films were grown on LSMO-buffered STO (001) substrates by pulsed laser deposition using a stoichiometric polycrystalline YHO target. The symmetric θ–2θ scan (Figure 1A) shows reflections corresponding to the LSMO (001) and YHO (111) planes, confirming that the film is oriented along the out-of-plane direction. The diffraction pattern exhibits the (001) reflection of the STO substrate, epitaxial LSMO (001) and YHO (111) layers at **2θ ≈ 22.76°**, **23.04°**, and **29.80°** respectively, confirming highly oriented growth of the heterostructure. No additional diffraction peaks are observed, indicating phase-purity.

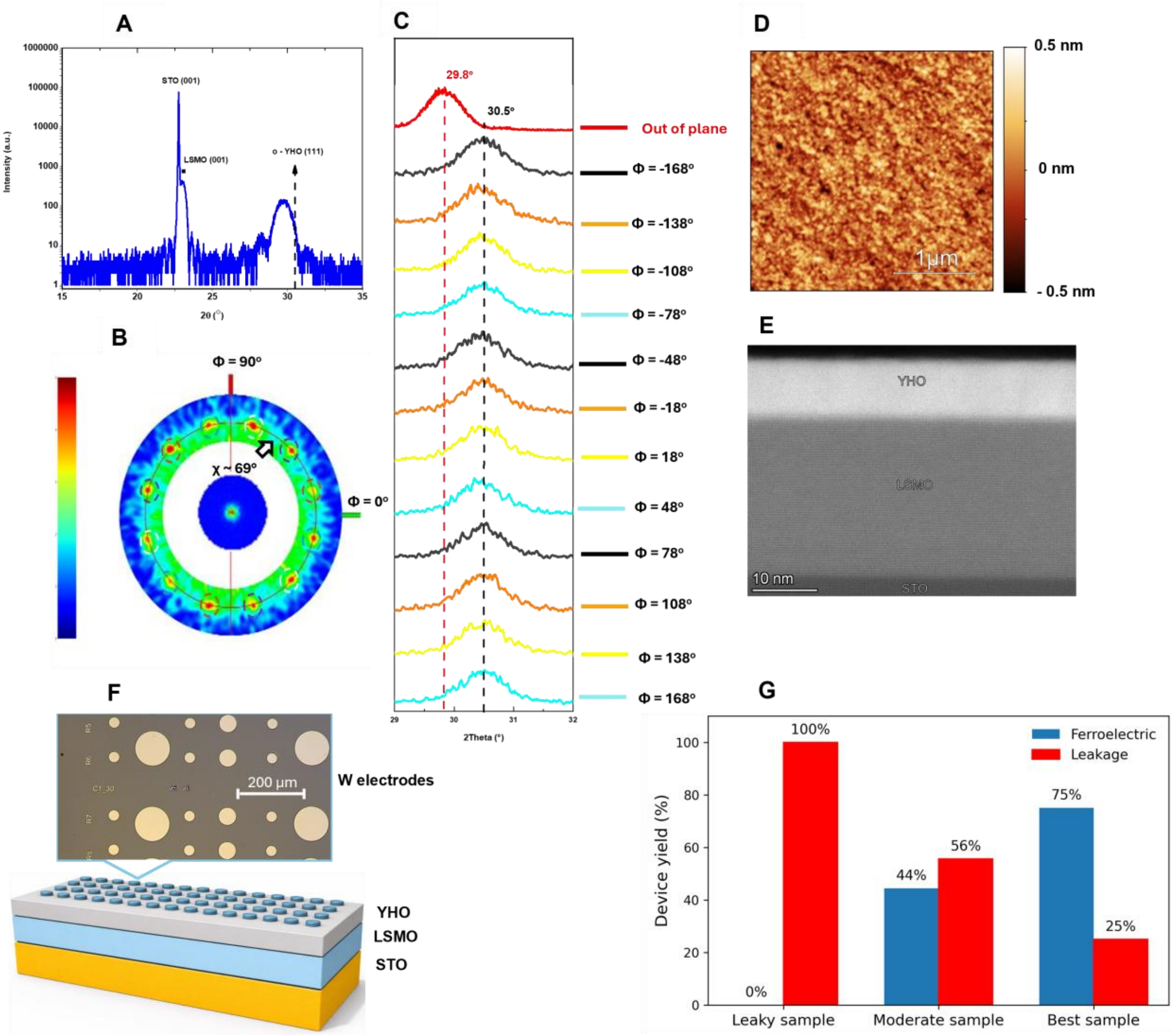

**Figure 1.** Structural, Morphological & Electrical characterization of epitaxial YHO thin films. **(A)** X-ray diffraction (XRD) θ–2θ scan of a 10 nm thick ferroelectric YHO film grown on an LSMO/STO heterostructure. **(B)** Pole figure around the (111) peak of a 10 nm YHO film at 2θ = 30.5°. The radial direction represents χ, while the azimuthal direction represents φ, with a (0°–360°) range. Colour represents intensity in log scale (left). **(C)** 2θ scans of the 13 peaks in the pole figure. **(D)** AFM topography image (3 μm × 3 μm) showing a uniform surface morphology and smooth surface across the scanned area. **(E)** HAADF-STEM image of the YHO/LSMO//STO heterostructure showing a well-defined layered structure with atomically sharp and abrupt interfaces and uniform film thickness. **(F)** Schematic of the capacitor structure. Circular W top electrodes are patterned on the YHO film grown on an LSMO bottom electrode on an STO substrate (W/YHO/LSMO/STO). **(G)** Device yield statistics comparing the fraction of ferroelectric and leaky capacitors across leaky, moderate, and best-performing samples.

To further determine the crystallographic phase of the YHO film, pole figure measurements were carried out by recording the {111} family of reflections (2θ ≈ 30.5°). Figure 1B shows twelve distinct poles, located at a tilt angle χ ≈ 69°, and one pole at χ ≈ 0°, consistent with (111) oriented films with four 90° in-plane rotated domain variants. Notably, all twelve poles appear at a higher 2θ value of ~30.5°, in contrast to the out-of-plane (111) reflection observed at ~29.80° (Figure 1C). This corresponds to 3 to 1 multiplicity of {111} family of planes where the out of plane d (111) (2.98 Å) is larger than the three inclined {111} family of planes (2.93 Å). This is consistent with rhombohedral symmetry as previously been reported for epitaxial rhombohedral $Hf_{0.5}Zr_{0.5}O_2$ films grown on LSMO-buffered $SrTiO_3$ substrates[10,11]. The AFM topography image (Figure 1D) reveals a smooth and homogeneous surface morphology with a low RMS roughness of 2.4 Å, indicating high-quality growth of the YHO film. HAADF-STEM analysis (Figure 1E) reveals a well-defined layered structure with clear Z-contrast between the YHO film, LSMO electrode, and STO substrate. The interfaces are sharp and abrupt, with no visible interdiffusion or secondary phases.

Ferroelectric capacitors were fabricated with tungsten circular electrodes patterned on YHO/LSMO//STO using standard photolithography followed by sputtering (see Figure 1F). Ferroelectric hysteresis was measured across multiple capacitors on films processed under identical optimized conditions, and with the same structural features (Figure S1) to assess device-to-device variability. Polarization-electric field (P–E) hysteresis measurements reveal three different types of films/samples, namely strongly leaky, moderately leaky and ferroelectric samples.

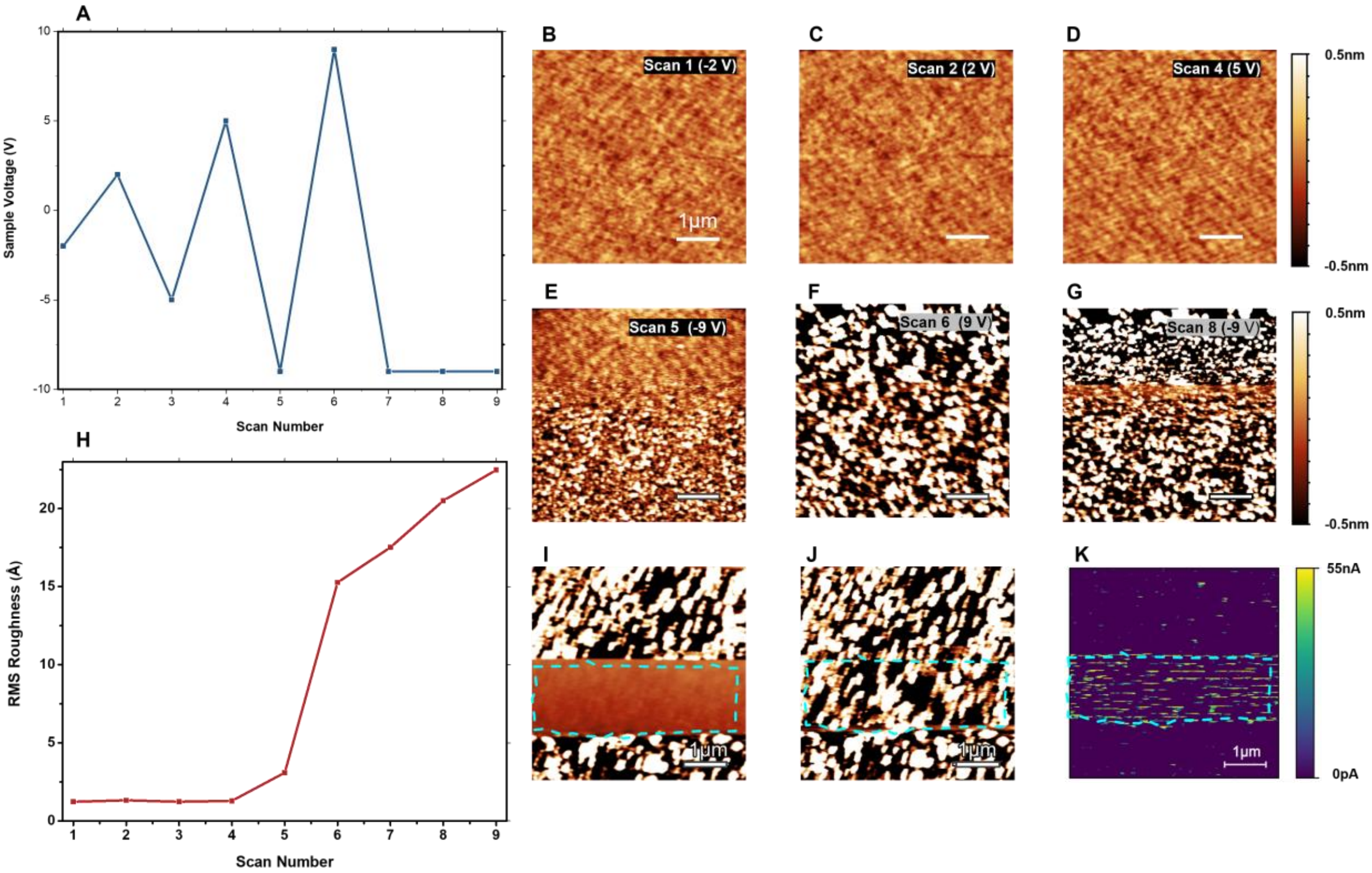

**Figure 2.** Surface roughness evolution with CAFM Scans for a leaky sample. **(A)**Variation of the applied sample bias with scan number during successive c-AFM scans. **(B–G)** AFM Topography Scans for different sample bias. **(H)** Roughness (R) as a function of scan number. **(I-K)** AFM topography images showing the surface morphology before and after scanning at the threshold voltage, together with the corresponding c-AFM conduction maps.

All the fabricated and tested capacitors (9) in the "leaky films" showed strong leakage behavior with rounded P-E loops (Figure S2A). In moderately leaky samples, ferroelectric behavior after wake-up was observed in 4 out of 9 capacitors (see Figure S2B for representative device PE loop), while the remaining devices remained leaky. In the best-performing films, 6 out of 8 capacitors exhibited clear ferroelectric switching, whereas the remaining two capacitors underwent dielectric breakdown before wakeup. $P_r$ in ferroelectric devices was extracted from PUND measurements (Figure S3) to be ~8.5 µC/cm². A summary of the device yield for the different samples is shown in Figure 1G.

**Surface roughness evolution with CAFM Scans for a leaky sample.**

Next, we investigate the evolution of the surface topography by scanning the voltage-biased sample with a conductive AFM tip. Prior to these measurements, the local ferroelectric hysteresis response of the film was characterized to identify regions exhibiting distinct electrical behavior. A region adjacent to a leaky electrode was subsequently selected and repeatedly scanned using conductive atomic force microscopy (c-AFM) while applying different sample biases. The applied bias was systematically varied from one scan to the next, and the corresponding bias sequence is shown in Figure 2A. Sample bias is cycled between positive and negative voltage, with the amplitude of voltage increasing with increasing scan number. No significant change in surface roughness is observed during the initial scans up to scan number 5 in AFM topography images (Figure 2B-E)**.** However, upon applying a sample bias of 9 V (scan number 6), a pronounced increase in roughness is observed (Figure 2F-G), with the surface roughness rising sharply from ~2 Å to ~1.5 nm. The roughness continues to increase with subsequent scans under bias. The evolution of surface roughness with scan number is summarized in Figure 2H. To further probe the bias-induced surface modification, a selected region of the sample was scanned under a threshold sample bias to induce pronounced roughening, while a nearby region was intentionally left unscanned as shown in Figure 2I. Upon subsequently scanning the entire area at this threshold voltage, the previously unmodified region also undergoes similar morphological deformation. The corresponding conductive AFM (c-AFM) maps reveal that conductivity (green to yellow regions in Figure 2K, corresponding to 30-50nA current) is localized only within the regions undergoing active deformation during the scan, whereas regions that were roughened in earlier scans no longer exhibit measurable current. Interestingly, this reveals that electrical conduction is transient and "leakage" is spatially correlated (Figure 2I-K) with surface roughening but diminishes once its complete.

**Surface roughness evolution with CAFM Scans for a ferroelectric sample.**

A similar set of measurements was performed on a region near an electrode exhibiting ferroelectric behavior. The selected area was subjected to successive biased scans, with the applied sample voltage varied as a function of scan number, with a protocol similar to that used in the leaky sample. The corresponding bias applied during each scan is summarized in Figure 3A. In contrast to the leaky region, the surface roughness in this area shows only a weak dependence on the applied sample bias, remaining on the scale of a few angstroms Å even at higher voltages. The AFM topography images as shown in Figure 3(B-G) indicate only minor morphological changes with successive scans. The roughness as a function of scan number (Figure 3H) shows only a slight increase, from ~1.8 Å to ~3.5 Å. The corresponding conductive AFM maps show no measurable current (Figure S4). A direct comparison of the surface roughness evolution for leaky and ferroelectric regions is presented in Figure 3I, where the roughness is plotted as a function of scan number for both cases. The leaky region exhibits a pronounced increase in roughness, with a sharp rise observed beyond a certain scan number, transitioning from few ångströms to >1.5 nm. In contrast, the ferroelectric region shows only a modest increase in roughness, remaining confined to the ångström scale even with similar scanning protocol.

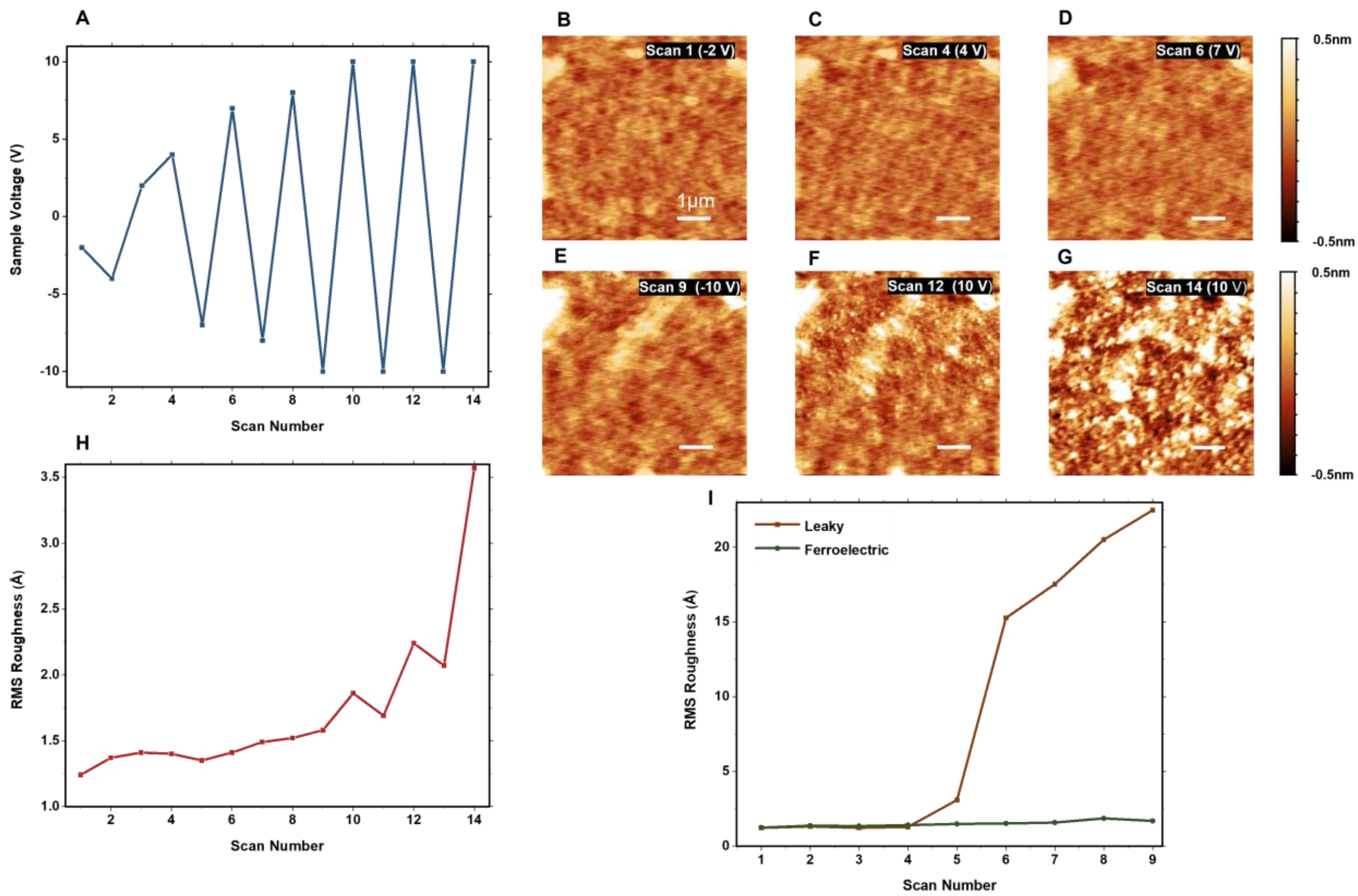

**Figure 3.** Surface roughness evolution with CAFM Scans for a ferroelectric sample. (**A**) Variation of the applied sample bias with scan number during successive c-AFM scans. **(B–G)** AFM Topography Scans for different sample bias. **(H)** Roughness (R) as a function of scan number for a ferroelectric sample. **(I)** Evolution of surface roughness as a function of scan number for leaky and ferroelectric samples obtained from AFM topography measurements.

## Local AFM and PFM contrast before and after electrical biasing of a ferroelectric sample.

Larger field of view AFM topography images reveal a clear height contrast between the biased, scanned and cycled region and the area surrounding it (Figure 4A), with the scanned area appearing elevated by ~2 nm relative to the pristine surface. For a similar scan-voltage profile as shown in Figure 4A, this height contrast develops progressively with successive scans (Figure 4B), with an initial sharp increase over the first four scans, followed by marginal increase over the subsequent scans (with error bars in Figure 4B reflecting the surface roughness).

PFM spectroscopy measurements were performed on both pristine and elevated (bias-cycled) regions of the sample. Figure 4C shows the amplitude and phase response when the bias is off.

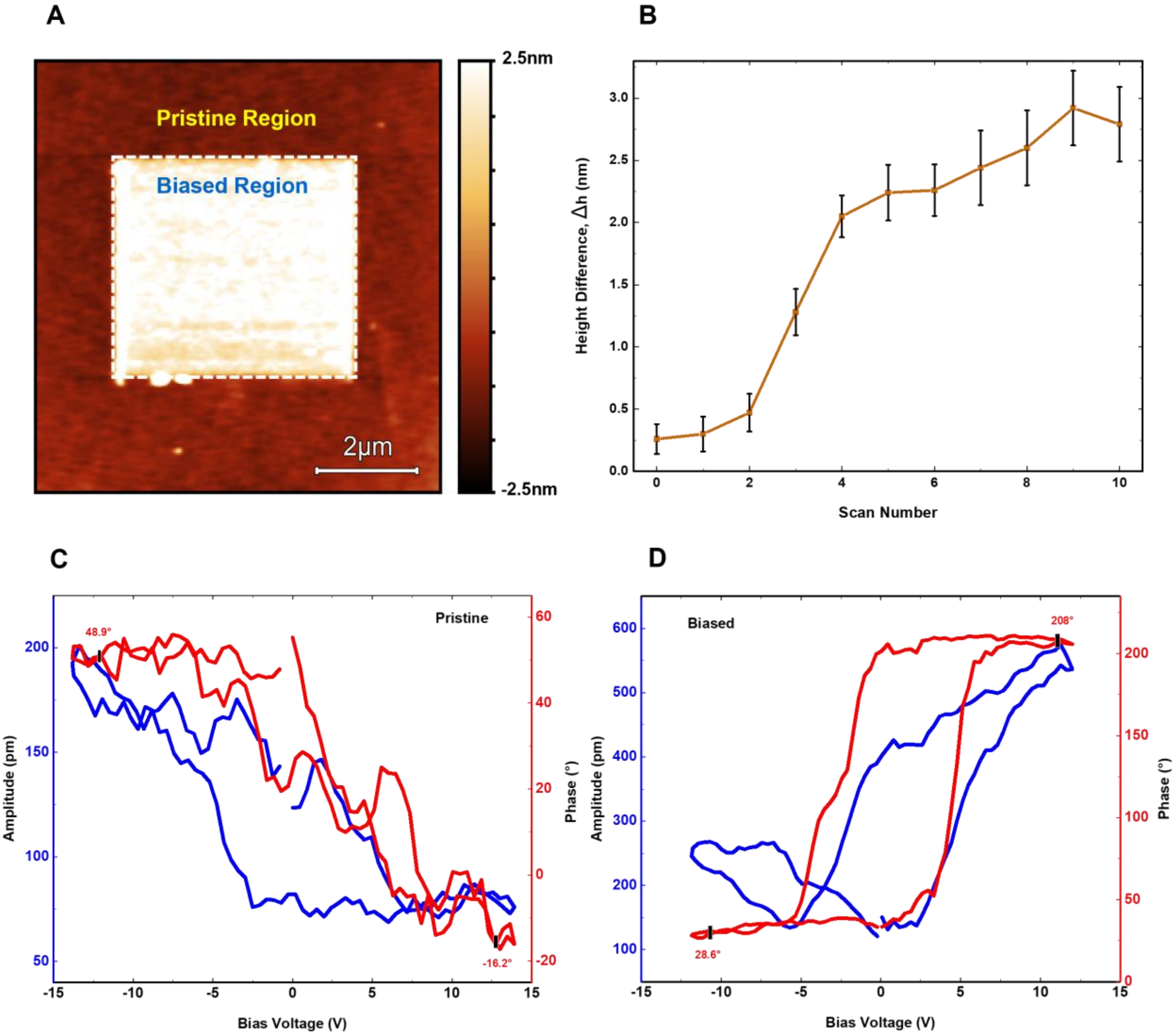

**Figure 4**. Bias-induced topographical evolution and local ferroelectric switching in epitaxial YHO thin films. (**A**) AFM topography image showing pristine and electrically biased regions. (**B**) Height contrast between pristine and biased region as function of scan number. (**C**) PFM hysteresis loops indicating local polarization switching in the pristine region. (**D**) PFM hysteresis loops indicating local polarization switching in the bias cycled region.

A characteristic butterfly loop and the phase signal showing no discernible switching. In contrast, measurements performed on the elevated region (bias-cycled region, Figure 4D) display well-defined ferroelectric switching, with clear butterfly-shaped amplitude loops and a phase hysteresis exhibiting a phase contrast close to ~180°, confirming the presence of switchable polarization. These results suggest that the increase in the height profile during the first few scans describe the topography changes associated with the wake-up effect. Notice that although epitaxial hafnia-based capacitors on LSMO/STO platform are described as wake-up free[8,10,11], in practice our samples still require a few cycles (weak wake up) or large initial voltage to trigger the ferroelectric behavior (Figure S5).

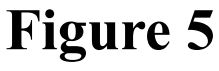

**Figure 5**

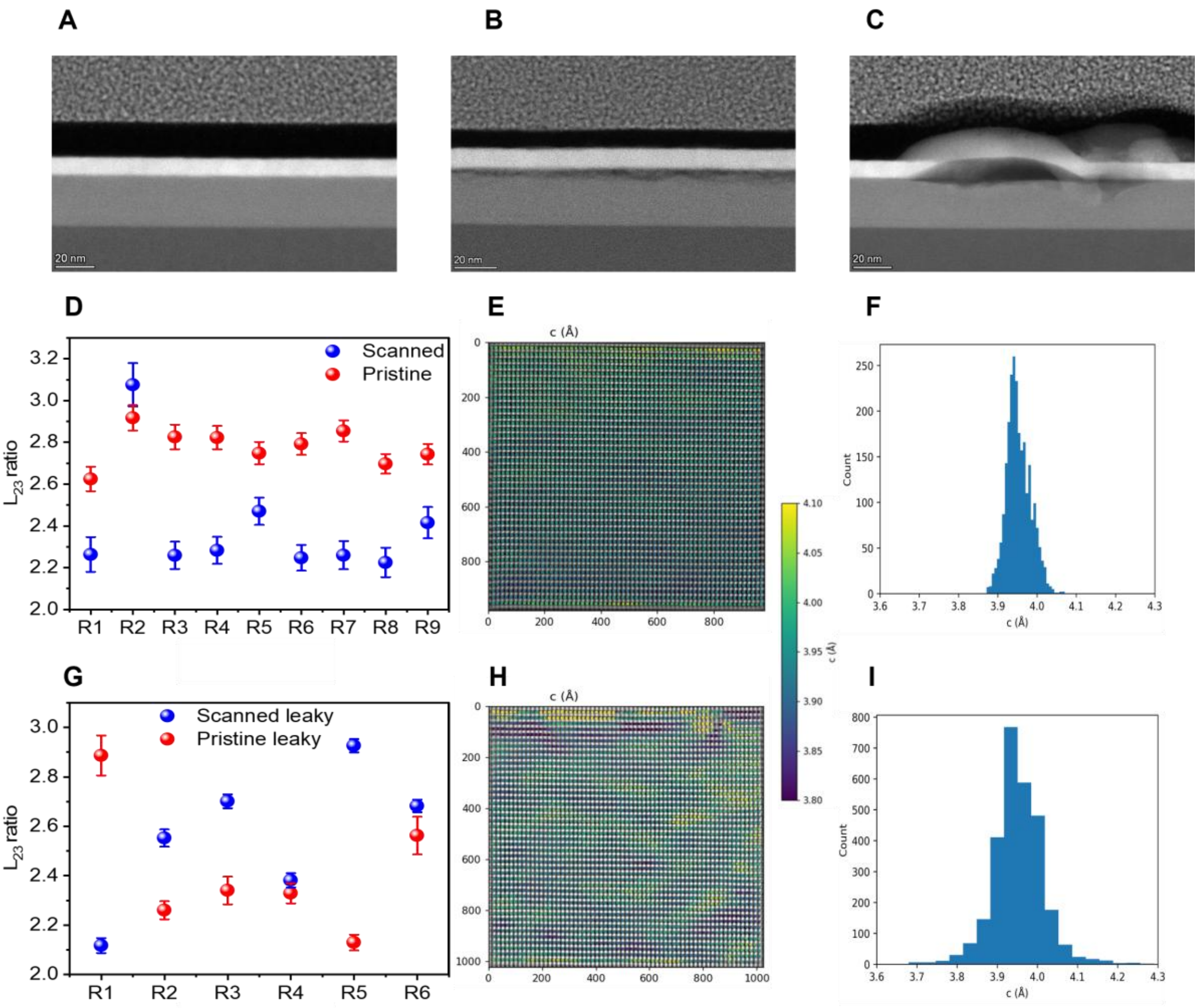

**Figure 5**. Spatial variation of the lattice parameter and Mn oxidation state in pristine, ferroelectric, and leaky regions of the LSMO bottom electrode. **(A–C)** Cross-sectional HAADF-STEM images showing the pristine region, electrically scanned ferroelectric region, and electrically scanned leaky region, respectively. **(D)** Mn $L_3/L_2$ intensity ratio measured at different regions before and after electrical scanning in the ferroelectric sample. **(E)** Spatial map of the out-of-plane lattice parameter (c) of the LSMO layer in the ferroelectric sample obtained from atomic-resolution STEM analysis. **(F)** Histogram showing the distribution of the c-lattice parameter extracted from panel **(E)**. **(G)** Mn $L_3/L_2$ intensity ratio measured at different regions in the pristine and electrically scanned leaky sample. **(H)** Spatial map of the LSMO c-lattice parameter in the leaky sample. **(I)** Corresponding histogram of the c-lattice parameter extracted from panel **(H)**.

Thus, while the leaky regions progressively roughen with cycling, the ferroelectric devices exhibit an initial increase in average height, consistent with the wake-up process, followed by a stabilization of the surface topography. Any subsequent increase in roughness at higher voltages and/or extended cycling can be attributed to the onset of leakage effects.

### STEM-EELS Analysis of Pristine and Bias-Cycled Regions

To understand the underlying electrochemical processes responsible for the bias-induced topographical evolution, cross-sectional STEM-EELS measurements were performed on electrically scanned ferroelectric (F) and leaky (L) regions together and compared with their pristine counterparts (PF and PL), as described in the Methods section. Representative cross-sectional HAADF-STEM images of the pristine, electrically scanned ferroelectric, and leaky regions are shown in Figure 5(A-C), respectively. The pristine region exhibits atomically sharp and abrupt interfaces between the YHO and LSMO layers (Figure 5A). Following wake-up, the interface remains continuous but develops a thin (~2 nm) darker contrast layer separating the epitaxial YHO and LSMO layers (Figures 5B, S6). This interfacial contrast correlates well with the bias-induced topographical evolution observed by AFM, where the electrically scanned region exhibits an approximately 2 nm increase in height (Figure 4A, B). In contrast, the electrically scanned leaky region exhibits pronounced structural degradation. The HAADF-STEM image (Figure 5C) reveals localized interfacial dark contrast together with blister formation within the YHO layer, indicating electrochemical reactions and gas evolution at these interfaces. This is consistent with the significant roughening observed in the AFM topography (Figure 2E-G). In addition, EDS analysis shows that the blistered YHO also contains signals of Mn and Sr (Figure S7), an effect of Joule heating due to conduction.

Mn $L_{2,3}$ edges were captured through STEM-EELS on the LSMO layers in the pristine and woken up regions. $L_3/L_2$ white line intensity ratio, estimated using a procedure described previously[37], reduces from $3.45 \pm 0.22$ to $2.91 \pm 0.07$, from pristine (PF samples) to woken up regions (F samples), consistent with the wake-up process resulting in oxidation of LSMO. Furthermore, measurements performed over multiple regions reveal a relatively uniform Mn $L_3/L_2$ white-line intensity ratio in both pristine and electrically scanned ferroelectric samples (Figure 5D), indicating homogeneous oxidation across the scanned area. A consequence of homogeneous oxidation state is also a narrow distribution of the out-of-plane lattice parameter (Standard Deviation, SD: 0.03 Å) of LSMO about 3.92 Å (Figure 5E-F).

On PL samples $L_3/L_2$ ratios across various regions shows variability (Figure 5 G), suggesting that Mn oxidation state is inhomogeneous in the sample, despite LSMO being epitaxial and single phase (Figure S1). Similar variability is carried over to the scanned leaky (L) samples (Figure 5G). Oxidation state inhomogeneity in LSMO is also reflected in wider variability (SD: 0.067 Å) and an inhomogenous spread of the out-of-plane lattice parameter about 3.92 Å (Figure 5H, I). Therefore, the leakage behavior and roughening transitions are fundamentally related to inhomogeneous oxygen vacancy distribution in LSMO.

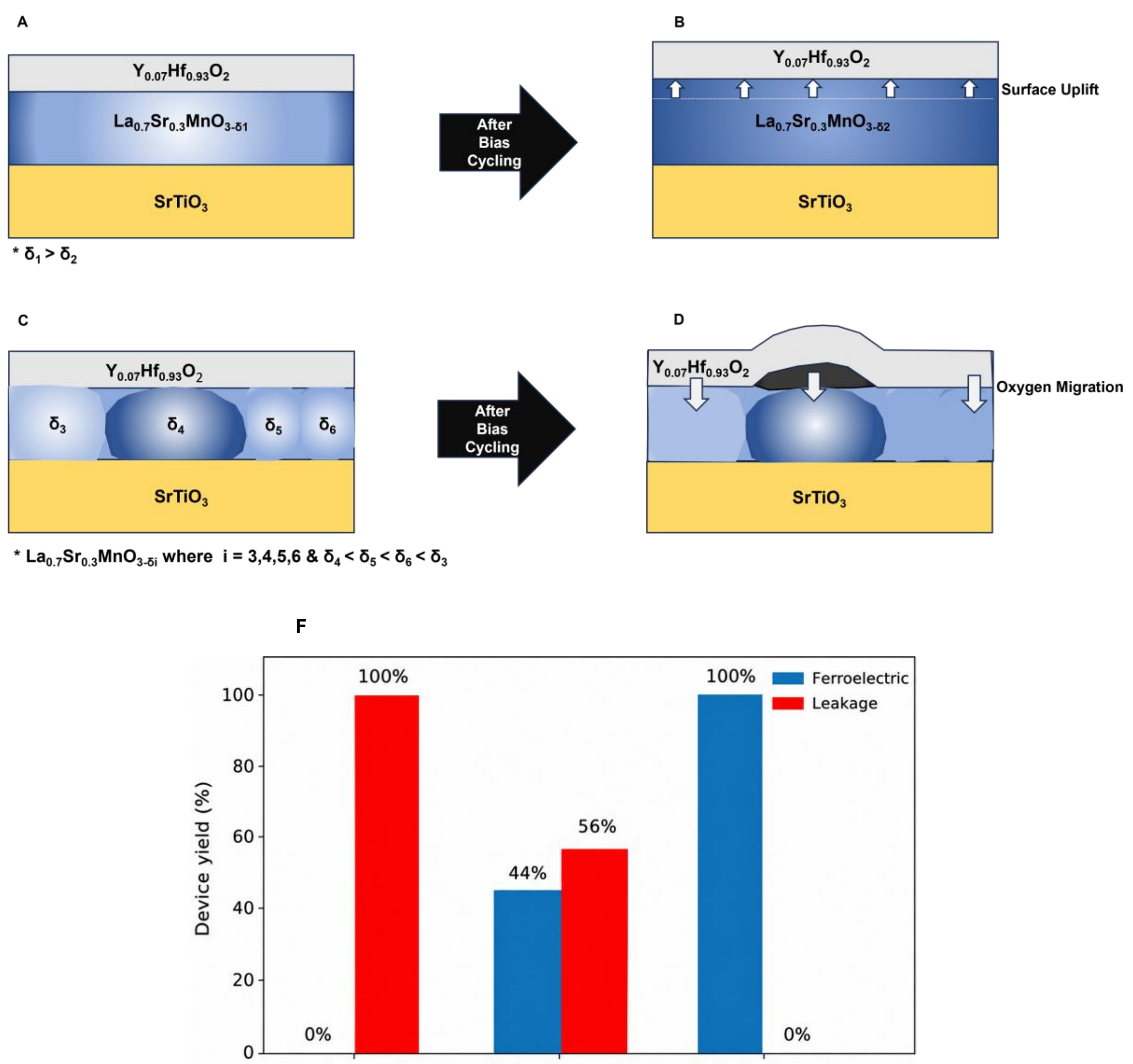

**Figure 6.** Proposed mechanism of oxygen-vacancy redistribution and topographical evolution in ferroelectric and leaky YHO films during electrical biasing. **(A)** Schematic of the pristine ferroelectric YHO/LSMO/STO heterostructure with a relatively uniform oxygen vacancy distribution in the LSMO bottom electrode. **(B)** After electrical bias cycling, the ferroelectric region exhibits a uniform surface uplift arising from the reversible redistribution of oxygen vacancies, accompanied by a reduced oxygen deficiency in the LSMO layer ($\delta_1 > \delta_2$). **(C)** Schematic of the pristine leaky region showing a non-uniform oxygen vacancy distribution in the LSMO bottom electrode ($\delta_4 < \delta_5 < \delta_6 < \delta_3$). **(D)** Following electrical bias cycling, localized oxygen migration promotes electrochemical reactions, resulting in oxygen gas evolution, blister formation, and irreversible surface deformation in the leaky region. **(F)** Device yield statistics comparing the fraction of ferroelectric and leaky capacitors across leaky, moderate, and optimized samples which incorporated feedback from operando SPM and STEM-EELS.

## Discussion:

YHO and LSMO form a coupled redox system, wherein the ferroelectric switching of YHO is mediated by oxygen exchange across the interface[25,34]. Our as grown LSMO is oxygen deficient. On samples where oxygen vacancies are homogenously distributed in LSMO (Figure 6A), wake-up involves uniform oxidation of the LSMO electrode or, equivalently, a uniform transfer of oxygen vacancies from LSMO into YHO. This process leads to oxidation of the LSMO layer, manifested experimentally as a uniform increase in the height of the scanned region (Figure 6B). Thus, LSMO acts as a reservoir of oxygen vacancies that can be reversibly exchanged with YHO during ferroelectric switching [25,34].

It is worth noting that the leaky samples also have LSMO layer where oxygen vacancies are inhomogeneously distributed (Figure 6C). Thus, in an oxygen exchange, regions with more oxygen content will not be able to accept oxygen from YHO layer, whereas regions with larger oxygen vacancy concentration can still accept oxygen, explaining the device-to-device variability. When LSMO ceases to be an oxygen reservoir, a competing interfacial reaction mechanism arises, resulting in evolution of molecular oxygen according to:

$$\mathbf{O_O^x} \rightarrow \frac{1}{2}\mathbf{O_2} + \mathbf{V_O^{\cdot\cdot}} + \mathbf{2}\boldsymbol{e}'$$

This oxygen evolution results in blistering at the LSMO-YHO interface (Figure 6D) observed in HAADF-STEM images (Figure 5C), and subsequent nanometric surface roughening as shown in AFM (Figure 2E-G) images. The accompanying release of electrons from this reaction contributes directly to the measured leakage current (see c-AFM data in Figure 2K). Notably, electron conduction is transient and spatially correlated with the roughening process. Thus conduction/leakage is observed only during active morphological evolution, i.e., while oxygen is being released, and the system reverts to an insulating state once the morphology stabilizes (regions outside the blue box) (Figure 2K).

These observations suggest that suppressing leakage and enabling robust ferroelectric switching (post wake-up) requires complete suppression of the OER pathway in favor of homogenous, oxygen migration across the YHO–LSMO interface. In our system, this requires LSMO to have a homogenous concentration of oxygen vacancies during processing. We leveraged these mechanistic insights to rationally engineer the oxygen stoichiometry of the YHO/LSMO heterostructure through systematic optimization of the post-deposition cooling profile, including both the oxygen pressure (to 10mbar) and cooling rate (to 6 °C $min^{-1}$). The oxygen stoichiometry of the heterostructure was improved, resulting in a more homogeneous Mn oxidation state in the LSMO layer. This optimization led 100% yield (10 tested) of ferroelectric capacitors (Figure 6F) on a film.

## Conclusion

We demonstrate that surface topography can serve as an effective local probe to understand leakage, ferroelectricity, and wake-up behavior in hafnia-based thin films. In the YHO/LSMO heterostructure, structurally similar samples exhibit markedly different topographic responses under biased scanning, highlighting the strong influence of local defect distribution and oxygen vacancy dynamics on the electrical behavior of the film.

Ferroelectric regions exhibit a gradual and spatially uniform elevation of the surface accompanied by only minimal changes in roughness, suggesting a homogeneous oxidation of the oxygen-deficient LSMO layer during wake-up cycling. In contrast, leaky regions show abrupt roughening and blister-like surface deformation under biased scanning. This behavior is attributed to rapid electrochemically driven oxygen evolution arising from non-uniform local oxidation states of Mn in the LSMO layer. The associated reaction simultaneously generates transient electronic conduction, which is observed only during active surface deformation and disappears once the structural modification is completed.

Using these observations as feedback, we optimized the film growth conditions to suppress leakage by homogenizing the oxidation state of Mn in the LSMO layer through a post-deposition oxygen annealing treatment. This improved control over the oxygen vacancy distribution enabled the stabilization of more uniform ferroelectric behavior across the films with significantly enhanced the device yield.

These observations establish a direct correlation between oxygen vacancy redistribution, local electrochemical processes, and the emergence of either ferroelectric or leakage-dominated behavior in hafnia thin films. The present work highlights the importance of nanoscale surface-probe combined with local spectroscopy in understanding defect-mediated switching phenomena and provides insight into controlling leakage and wake-up processes in hafnia ferroelectrics.

## Experimental / Method

### Thin-film synthesis

Epitaxial YHO thin films were deposited on LSMO-buffered (001) $SrTiO_3$ substrates grown using pulsed laser deposition (PLD) with a wavelength of 248nm. The base pressure of the PLD chamber is about 2.8 $e^{-7}$ mbar. LSMO layer with a thickness of ~30nm was deposited using a laser fluence of 1.5 J $cm^{-2}$ at a laser frequency of 1 Hz under a 0.15 mbar oxygen atmosphere and a substrate temperature of 750 °C. A ceramic 7% Y-doped $HfO_2$ target was synthesized at 1,400 °C by solid-state reaction, starting from 99.99% $Y_2O_3$ and 99.95% $HfO_2$ powders.

YHO films of thickness ranging from 7-10nm were deposited using a laser fluence of 1.8 J $cm^{-2}$ and a repetition rate of 2 Hz. The chamber was maintained at an oxygen pressure of 0.1 mbar, while the substrate temperature was held at 750°C. Following deposition, the samples were cooled to room temperature at a rate of 6 °C $min^{-1}$ under an oxygen pressure of 10 mbar.

### X-ray structural characterization

Structural characterization was performed using X-ray diffraction techniques, including $\theta$–$2\theta$ scans, rocking curve measurements, and X-ray reflectivity, carried out on a Rigaku Smart Lab diffractometer employing Cu K$\alpha$ radiation ($\lambda$ = 1.54 Å). Phase confirmation was achieved through pole figure measurements performed using $\varphi$ scans, varying the chi while the detector is at a fixed Bragg reflection.

### Hysteresis measurements

Tungsten top electrodes with a thickness of approximately 50 nm were deposited ex situ by sputtering at room temperature under vacuum conditions. Ferroelectric measurements were performed at room temperature by applying voltage pulses to the W top electrodes, with diameters ranging from 30 to 150 µm, while the bottom LSMO electrode was grounded. The measurements were carried out using a Radiant Technologies Precision Multiferroic ferroelectric tester.

### Scanning probe microscopy

AFM measurements were performed using a Park AFM system with TESPA V2 probes. PFM and conductive AFM (c-AFM) measurements were carried out using a commercial atomic force microscopy system (MFP-3D, Asylum Research) equipped with Pt-coated probes (Budget Sensors EZ-75). PFM measurements were conducted using an a.c. modulation voltage of 0.65 V at a resonance frequency of approximately 380 kHz applied along with the DC bias applied to the tip. During c-AFM measurements, a DC bias was applied to the sample through the grounded conductive AFM tip, while the LSMO bottom electrode served as the electrical contact.

**Electron Microscopy**

Scanning transmission electron microscopy (STEM) imaging was performed using a high-angle annular dark-field (HAADF) detector with a collection angle ranging from 48 to 200 m rad and a convergence semi-angle of 24 m rad. All imaging was carried out at an accelerating voltage of 300 kV with a probe current of approximately 20 pA using an aberration-corrected Thermo Fisher Scientific Titan Themis 300 microscope. Energy-dispersive X-ray spectroscopy (EDS) was conducted using a ChemiSTEM Super-X quad detector system, with extended acquisition times employed to obtain sufficient signal for quantitative analysis.

Electron energy-loss spectroscopy (EELS) measurements were performed at an accelerating voltage of 300 kV using a dispersion of 100 meV, without a monochromator. Two-dimensional spectrum imaging was acquired to improve signal-to-noise ratio for subsequent analysis. Following acquisition, background subtraction was carried out using a power-law fit. The Mn $L_{2,3}$ white-line edges were analyzed using appropriate Hartree–Slater cross sections, including corrections for plural scattering. A consistent energy window of approximately 80 eV was used throughout the analysis. The processed spectra were subsequently fitted with Gaussian functions to extract peak areas, which were used to determine the Mn $L_3/L_2$ intensity ratio.

# AUTHOR INFORMATION

## Corresponding Author

Pavan Nukala
Centre for Nano Science and Engineering,
Indian Institute of Science,
Bengaluru 560012, Karnataka, India
Email: pnukala@iisc.ac.in

Anudeep Tullibilli
Centre for Nano Science and Engineering,
Indian Institute of Science,
Bengaluru 560012, Karnataka, India
Email: anudeept@iisc.ac.in

## Author Contributions

AT, PN conceived the ideas, co-wrote the manuscript. AT performed all the experiments including synthesis, XRD, SPM and analysed the data. KB and SP performed STEM-EELS and analysis of corresponding data. All authors discussed the results and approved the final version of the manuscript.

## Acknowledgements:

The authors acknowledge the Micro and Nano Characterization Facility (MNCF) and the National Nanofabrication Centre (NNfC) at the Centre for Nano Science and Engineering (CeNSE), Advanced Facility for Microscopy and Microanalysis (AFMM), Indian Institute of Science (IISc), Bengaluru, for access to fabrication and characterization facilities. This work was supported by funding from SERB (DST), New Delhi, Government of India (CRG/2022/003506), as well as support from the DST-COE on piezoMEMS (DST/TDT/AM/2022/084).

**Declaration:** LLMs such as ChatGPT/Gemini was used solely to improve grammar, sentence structure, and language clarity where necessary. These were not used to generate scientific content, interpret data, perform analyses, or produce any results or conclusions.

## Supporting Information:

The Supporting Information includes Figures S1–S7.

# Supporting Information

## Operando Surface Probe Microscopy Reveals Electrochemical Origins of Reliability Variability in Hafnia Ferroelectrics

*Anudeep Tullibilli*, Kartick Biswas, Shubham Kumar Parate, and Pavan Nukala**

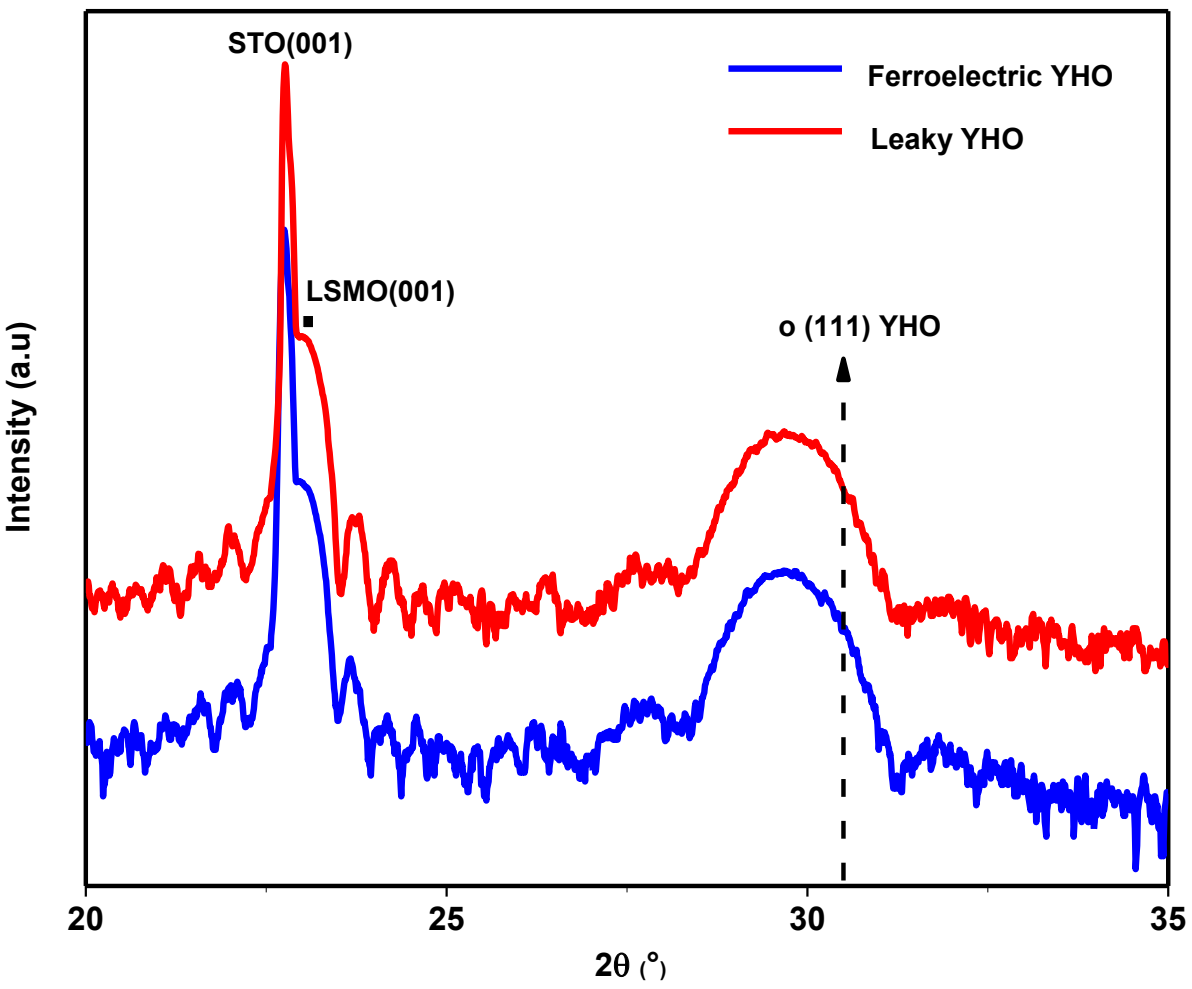

**Figure S1.** X-ray diffraction (XRD) θ–2θ scans of ferroelectric and leaky YHO thin films grown on LSMO-buffered STO (001) substrates. Both samples exhibit very similar XRD and are grown in a phase-pure (111) oriented rhombohedral phase (reflection near 2θ ≈ 29.8 °)

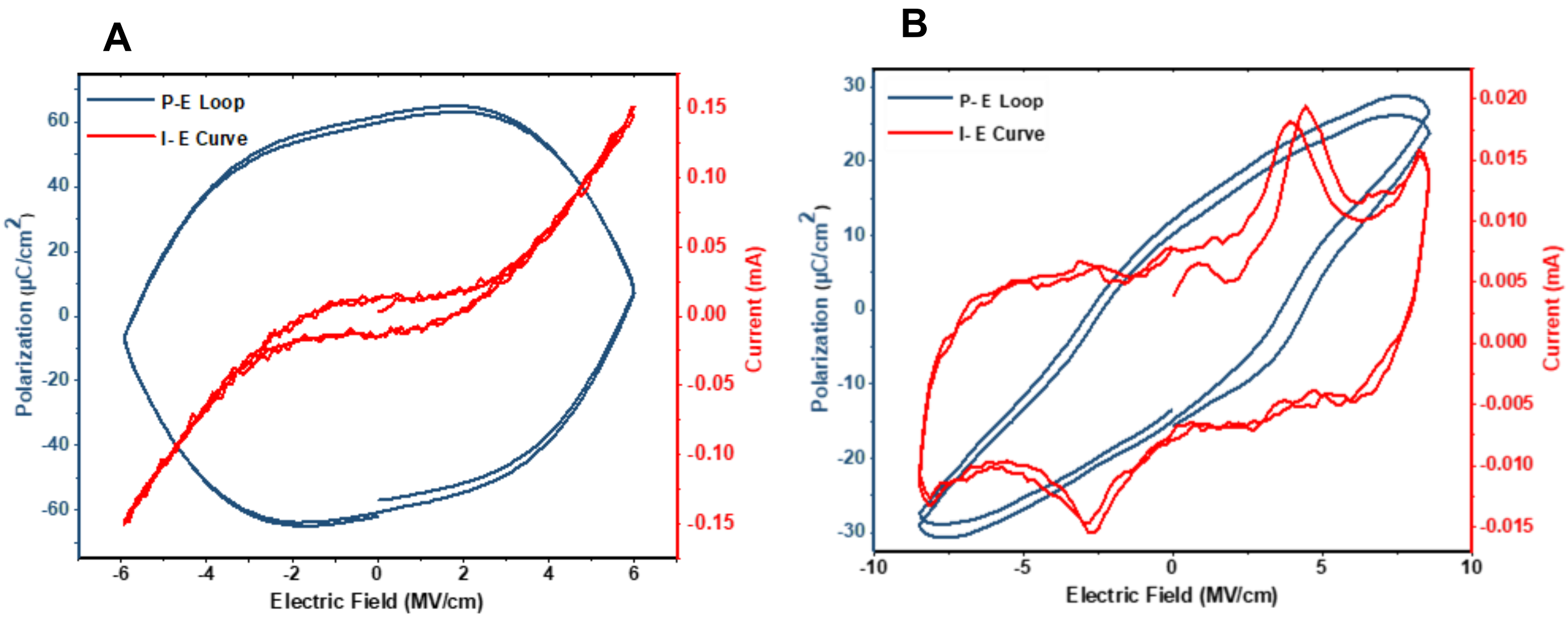

**Figure S2**: **A)** P–E (blue) and I–E (red) hysteresis of a leaky capacitor, showing a rounded polarization loop and monotonically increasing current due to leakage. **B)** P–E (blue) and I–E (red) hysteresis of a ferroelectric

capacitor after wake-up, showing a well-defined polarization loop with distinct switching current peaks near the coercive field.

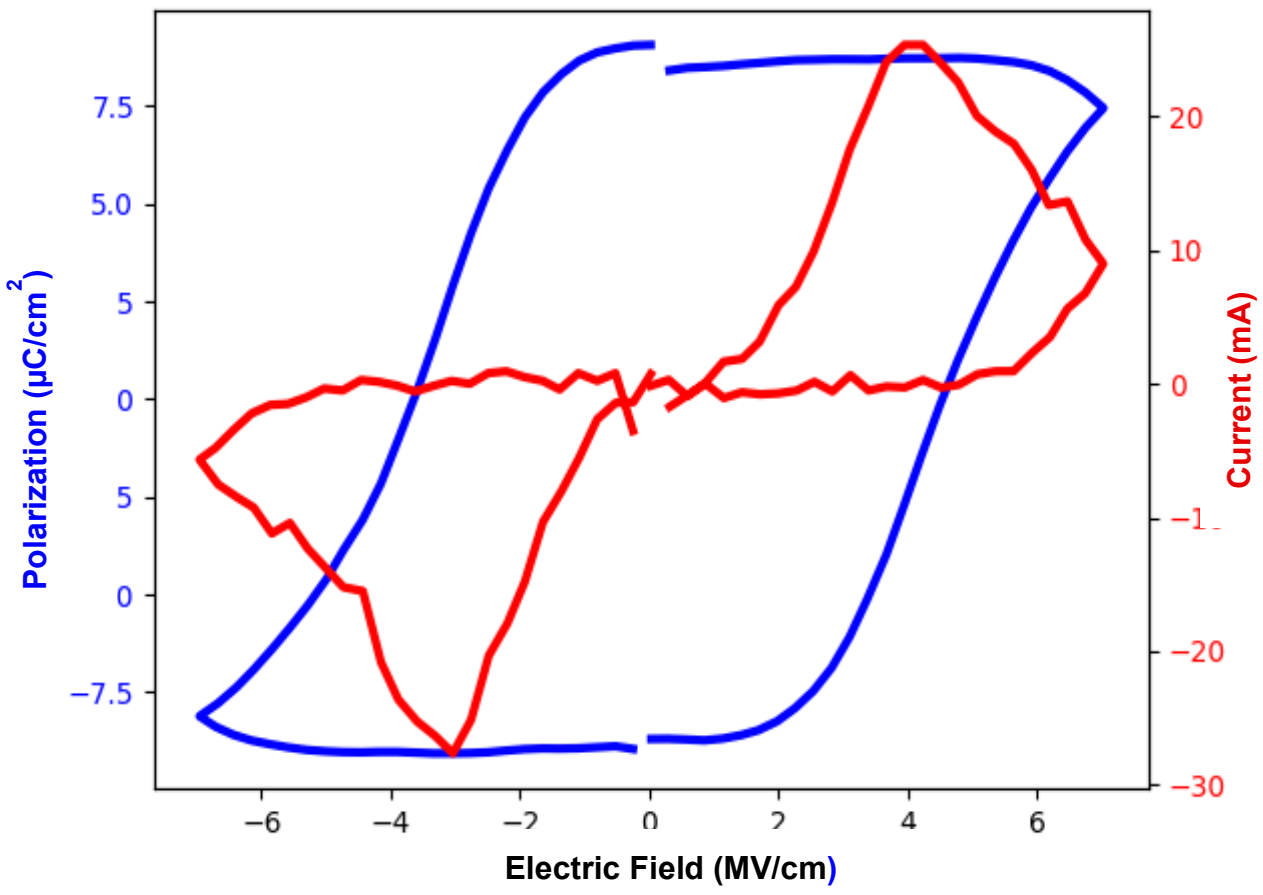

**Figure S3**. Positive-up negative-down (PUND) measurement of the epitaxial YHO capacitor showing the polarization–electric field (blue) and switching current–electric field (red) characteristics. The device exhibits a remanent polarization ($P_r$) of approximately **8.5 μC cm⁻²** with a coercive field (Ec) of approximately **3.5 MV cm⁻¹.**

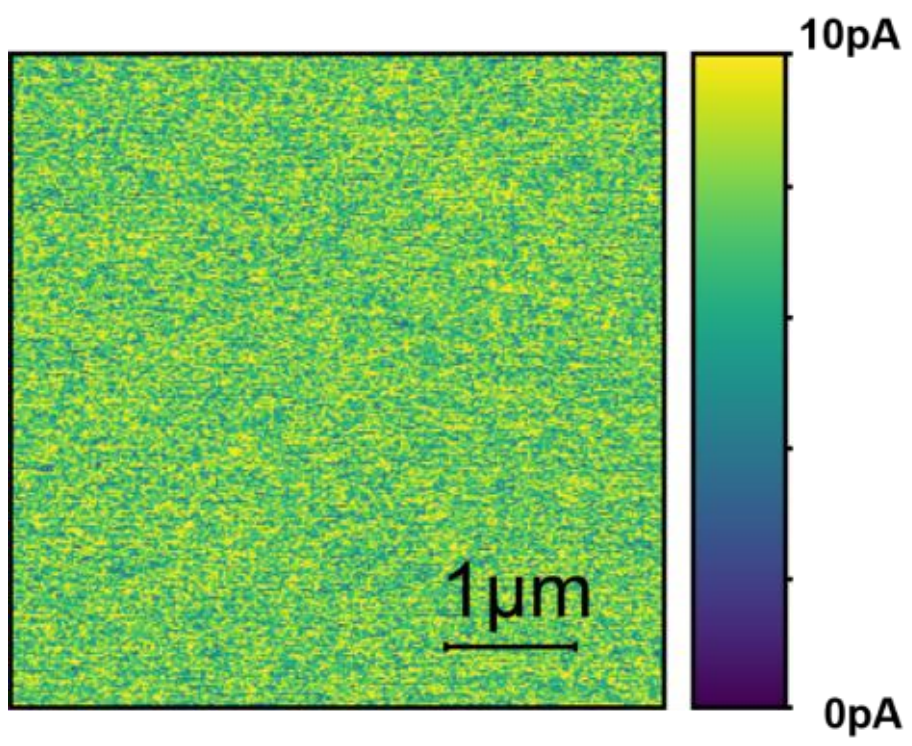

**Figure S4**. Conductive AFM (c-AFM) current map of the ferroelectric region of YHO thin film acquired under the measurement conditions described in the Methods. Leakage current measured in these ferroelectric samples <10 pA. In contrast leakage currents measured in local regions on the leaky samples are ~50 nA.

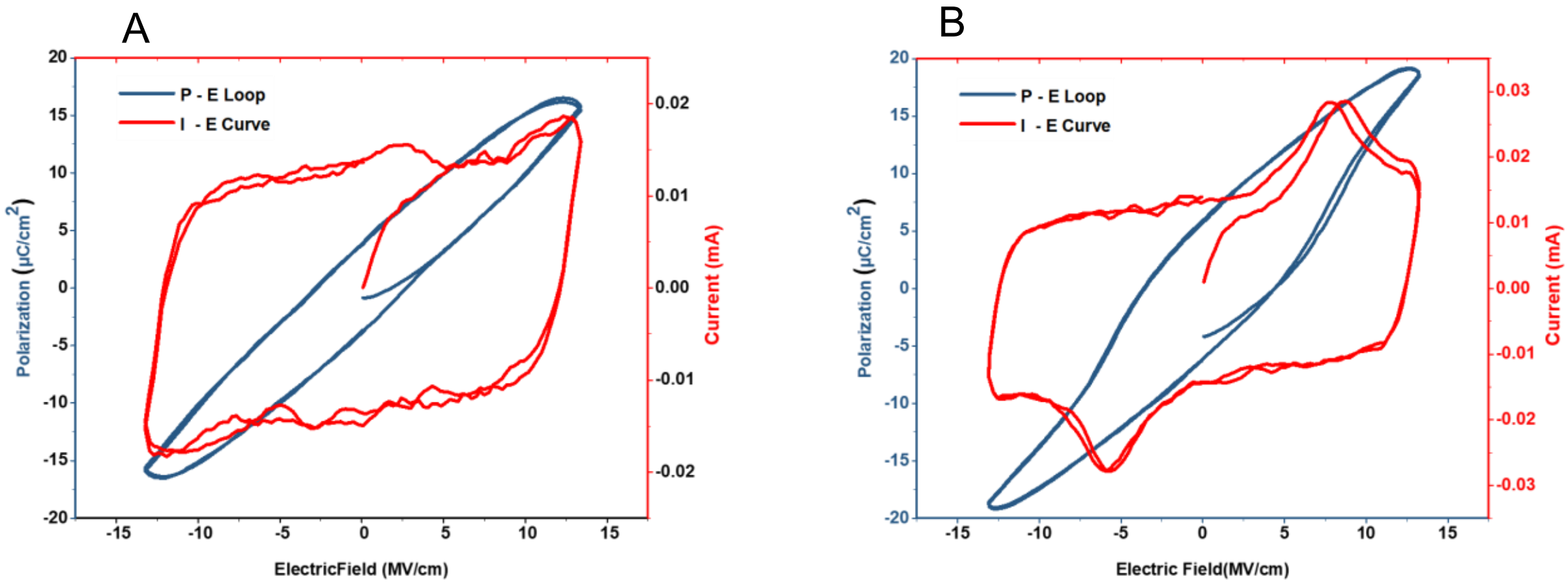

**Figure S5**. Polarization–electric field (P–E) hysteresis loops (blue) and corresponding current–electric field (I–E) characteristics (red) of the epitaxial YHO capacitor measured **(A)** before wake-up and **(B)** after wake-up. The wake-up state was achieved by a applying a higher electric field to the capacitor.

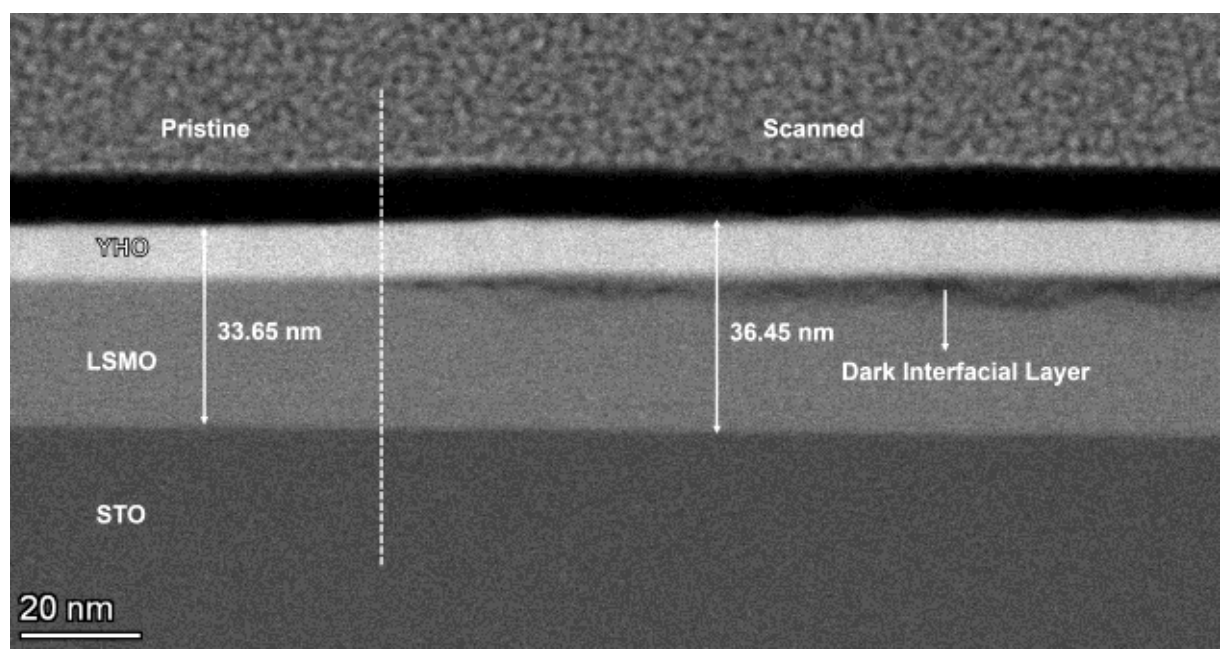

**Figure S6**. Cross-sectional HAADF-STEM image comparing the pristine and electrically scanned (woken-up) regions of the YHO/LSMO/STO heterostructure. The electrically scanned region exhibits an increase in the YHO/LSMO layer film thickness from 33.65 nm to 36.45 nm, together with the appearance of a thin dark interfacial layer (~2 nm) between the YHO and LSMO layers.

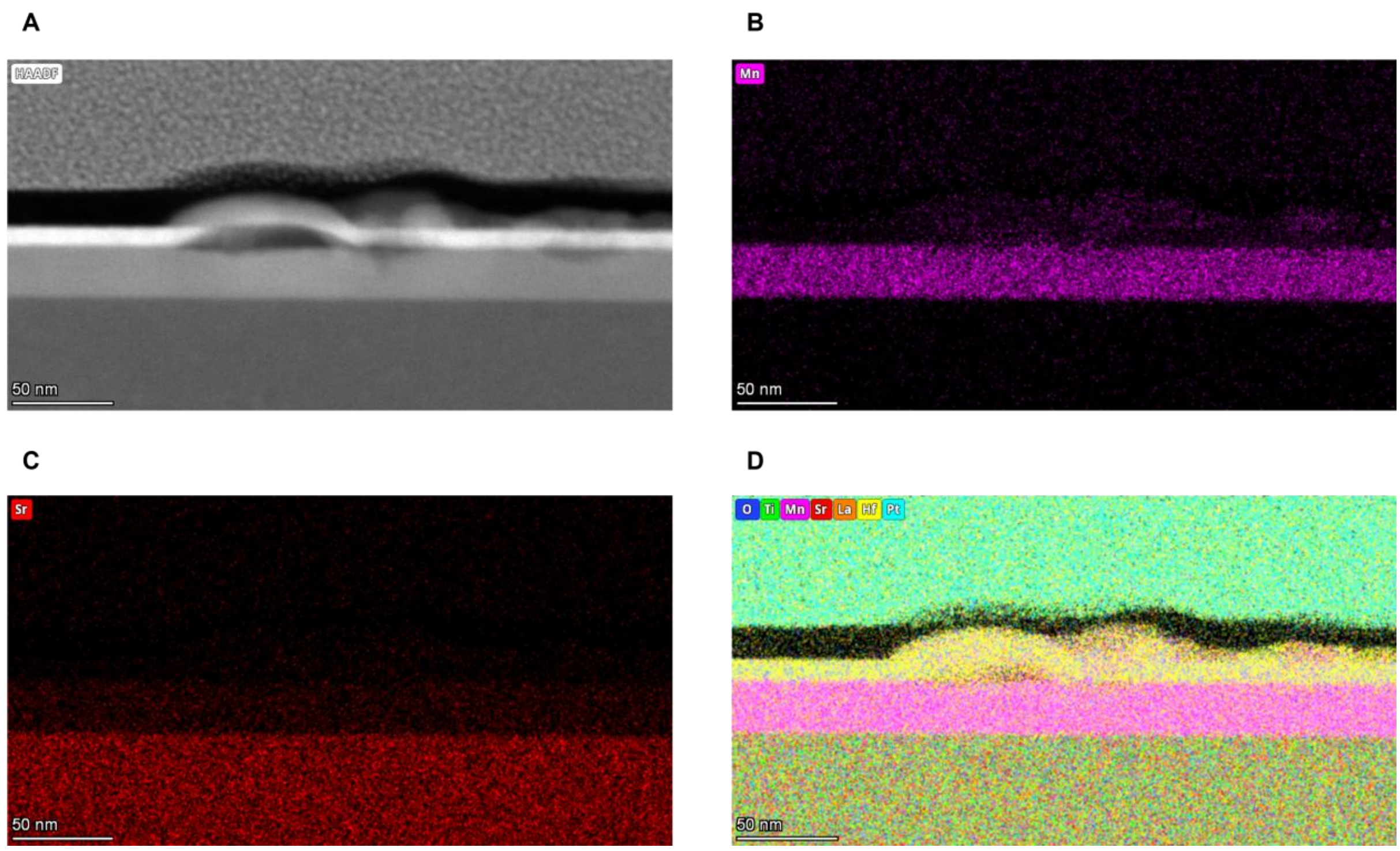

**Figure S7**. STEM-EDS analysis of the blistered region. (**A**) HAADF-STEM image of the blistered YHO/LSMO interface. (**B**) Mn elemental map. (**C**) Sr elemental map. (**D**) Composite EDS elemental map showing the spatial distribution of the constituent elements. The presence of Mn and Sr signals within the blistered region indicates elemental transport from the LSMO electrode into the YHO layer following electrical biasing.